\documentclass[a4paper]{article}
\usepackage[latin1]{inputenc}
\usepackage{graphicx}
\usepackage{url}

\begin{document}

\title{POLARIZED POSITRONS FOR THE ILC --- \\
UPDATE ON SIMULATIONS\thanks{Talk was presented at the POSIPOL 2011
Workshop, IHEP, Beijing, 28--30 August 2011.}}

\author{\vspace{2mm}A. USHAKOV, O. S. ADEYEMI and G. MOORTGAT-PICK \\
II. Institute for Theoretical Physics, University of Hamburg,\\
\vspace{5mm}Luruper Chaussee 149, D-22761 Hamburg, Germany\\
\vspace{2mm}F. STAUFENBIEL and S. RIEMANN \\
DESY Standort Zeuthen\\
Platanenallee 6, D-15738 Zeuthen, Germany}

\date{}

\maketitle{}
\vspace{-75mm}
\hfill DESY-12-018
\vspace{65mm}

\begin{abstract}
To achieve the extremely high luminosity for colliding
electron-positron beams at the future International Linear Collider \cite{RDR}
(ILC) an undulator-based source with about 230 meters helical
undulator and a thin titanium-alloy target rim rotated with
tangential velocity of about 100 meters per second are foreseen. The
very high density of heat deposited in the target has to be
analyzed carefully. The energy deposited by the photon beam in the
target has been calculated in FLUKA. The resulting stress in the
target material after one bunch train has been simulated in ANSYS.
\end{abstract}

%\keywords{ILC; positron source; thermal stress.}

%\bodymatter

\section{Introduction}

The positron-production target for the ILC positron source is driven
by a photon beam generated in an helical undulator placed at the end
of main electron linac~\cite{SB2009}. The undulator length is chosen to
provide the required positron
yield. The source is designed to deliver 50\% overhead of positrons.
Therefore, the positron yield has to be 1.5 positrons per electron
passing the undulator. The required active length of the undulator
is about 75~meters for the nominal electron energy of 250 GeV, the undulator
$K$-value has been chosen to be 0.92, the undulator period is 11.5 mm and a
quarter-wave transformer is used as optical matching device (OMD).
The photon first harmonic energy cutoff is 28 MeV, 
the average energy of photons is about 29 MeV and the average photon
beam power is about 180 kW in a train of 2625 bunches with a
frequency of 5 Hz. Although only relatively small fraction of
total photon beam energy deposited in the target (about~5\%), the peak
energy density deposited in target is high due to the small opening
angle of the synchrotron radiation in the helical undulator
resulting in a small photon spot size on the target. For example,
for 500~meters space between the undulator and target, the average
radius of the photon beam is approximately 2 mm and the peak energy
density could achieve 120 J/g in the 0.4 radiation length thick
titanium-alloy target rotated with 100~m/s tangential velocity.

There is no experimental data indicating the upper limit of the peak
energy density deposited by photons in the titanium alloy material
with 90\% of titanium, 6\% of aluminium and 4\% of vanadium. The
analysis of the electron beam induced damage to the SLC positron
target \cite {Bhar01} and the simulations of thermal shock \cite
{Stein01} show that the energy deposition limit is about 30~J/g for
tungsten with 25\% of rhenium target irradiated by 33~GeV electrons
and a general criteria of failure due to an equivalent (von-Mises)
stress of 50\% of tensile strength may apply to this target material
\cite{Stein01}.

The thermal structural modeling of a rotated titanium target
irradiated by helical undulator photons has been performed for the
NLC by W.~Stein and J.~Sheppard \cite{Stein02}. They recommend to
consider as ``safe'' thermal stresses below one third to one half
of the yield stress.

In this paper, the thermal stress in the ILC positron source target
has been estimated for the SB2009 set of parameters \cite{SB2009}.

\section{Energy Deposition and Temperature in \\Target.
Static Model of Material Response}

The energy transfer from the photon beam into temperature of target
material and the structural deformation and mechanical stress
coupled with this temperature rise due to complexity of these
time-dependent, cross-coupled and nonlinear processes cannot be
treated with the highest level of details \cite{Zaz95}. Therefore,
the choice of simulation tools and reasonable approximations and
simplifications plays an important role.

The energy deposition in the target has been calculated in FLUKA
\cite{FLUKA}. An amount of energy is counted as deposited if after
collisions the primary or secondary particles have energies lower
than the energy cut-offs. The FLUKA default cut-offs were used:
1.511 MeV -- for electrons and positrons and 333~keV -- for photons.

\begin{figure}[t]
\begin{center}
\includegraphics[width=2.4in,keepaspectratio=true]{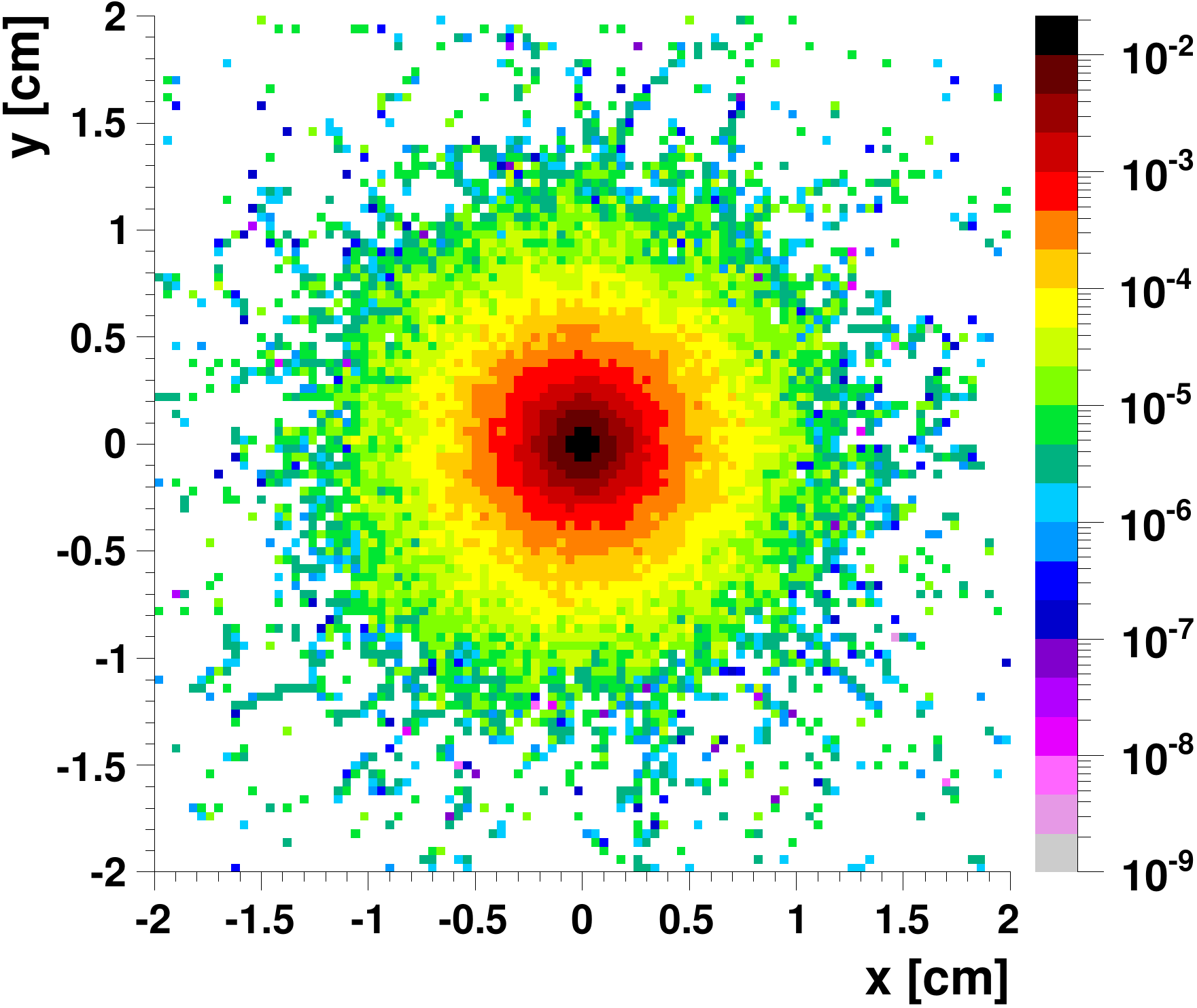}
\end{center}
\caption{Distribution (dependence in $x$ and $y$) of the energy
[GeV/(ph\,cm$^{3}$)] deposited close to the backside of Ti6Al4V
target calculated in FLUKA.}
\label{au:fig-Edep-fluka}
\end{figure}

Figure~\ref{au:fig-Edep-fluka} shows the ``original'' FLUKA data
distribution (i.e. without any scaling factors)  of the energy
deposited close to the back side of the target. The energy is given
in units of GeV per cubic centimeter and per impinging on the target
photon.

The temperature rise $\delta T$ in the target for given a energy
deposition $E_{dep}$ has been calculated according to the following
equation
\[
 \delta T = \frac{E_{dep} N_{e^{-}} Y_{ph} L_{und} N_{b}}{\rho c_{p}},
\]
where $N_{e^{-}}$ is the number of electrons per bunch ($2 \times
10^{10}$), $Y_{ph}$ is the photon yield (1.94 photons per electron and
per 1 meter of undulator), $L_{und}$ is the length of undulator (70
meters), $N_{b}$ is the number of bunches crossing the same
volume/bin, $\rho$ is the target density (4.49 g/cm$^{3}$) and $c_{p}$
is the specific heat capacity (0.523 J/(g K)).

\begin{figure}
\begin{center}
\includegraphics[width=3.5in,keepaspectratio=true]{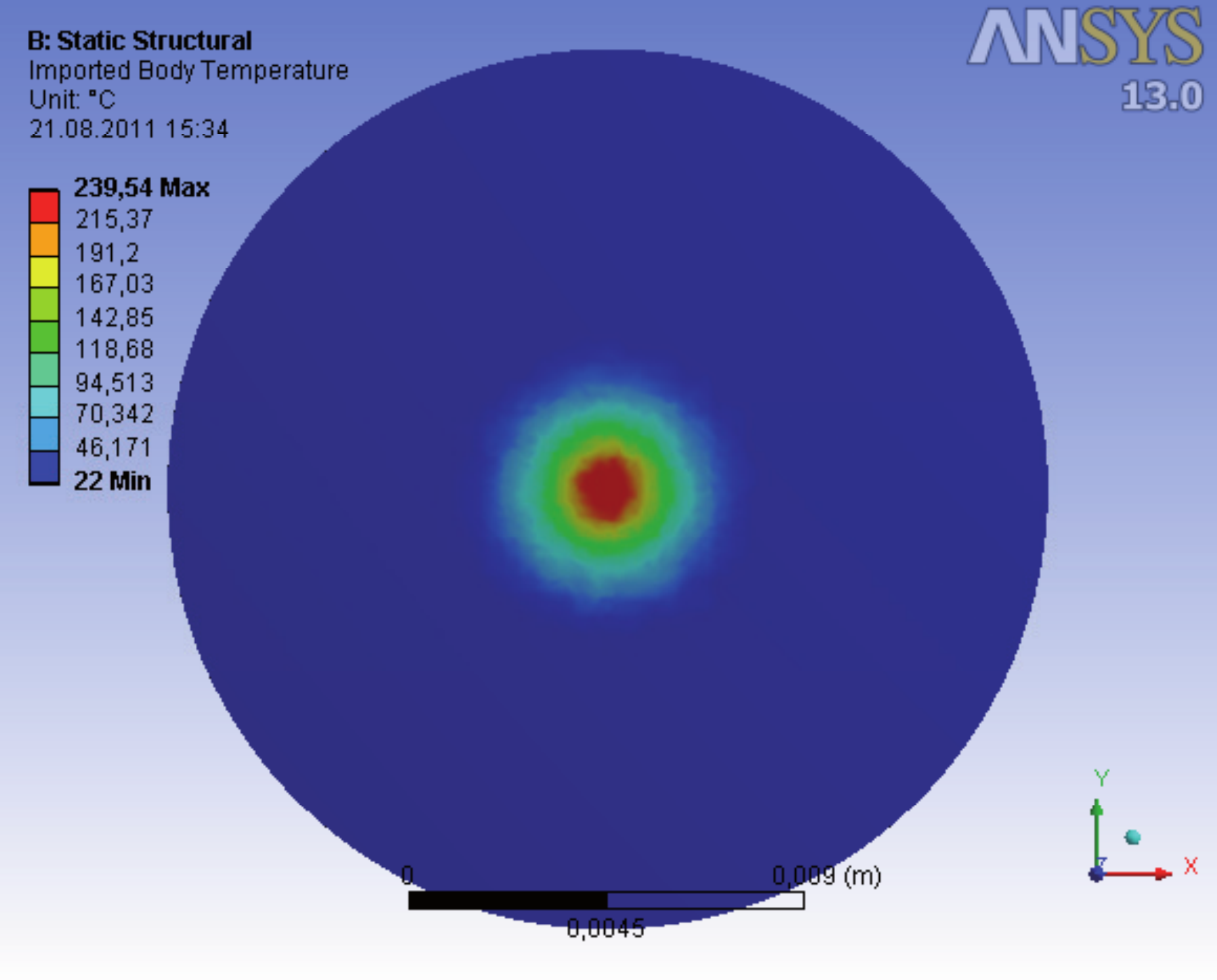}
\end{center}
\caption{Temperature profile on target back side after 100 bunches.}
\label{au:fig-Temperature-backside}
\end{figure}

The temperature data in a 1.48 cm thick cylindrical titanium target
after the first 100 bunches has been imported into ANSYS \cite
{ANSYS}. The temperature distribution on the back side of the target
is shown in Figure~\ref{au:fig-Temperature-backside}. The maximal
increase of temperature per bunch is about 2.2~K.

\begin{figure}
\begin{center}
\includegraphics[height=2.35in,keepaspectratio=true]{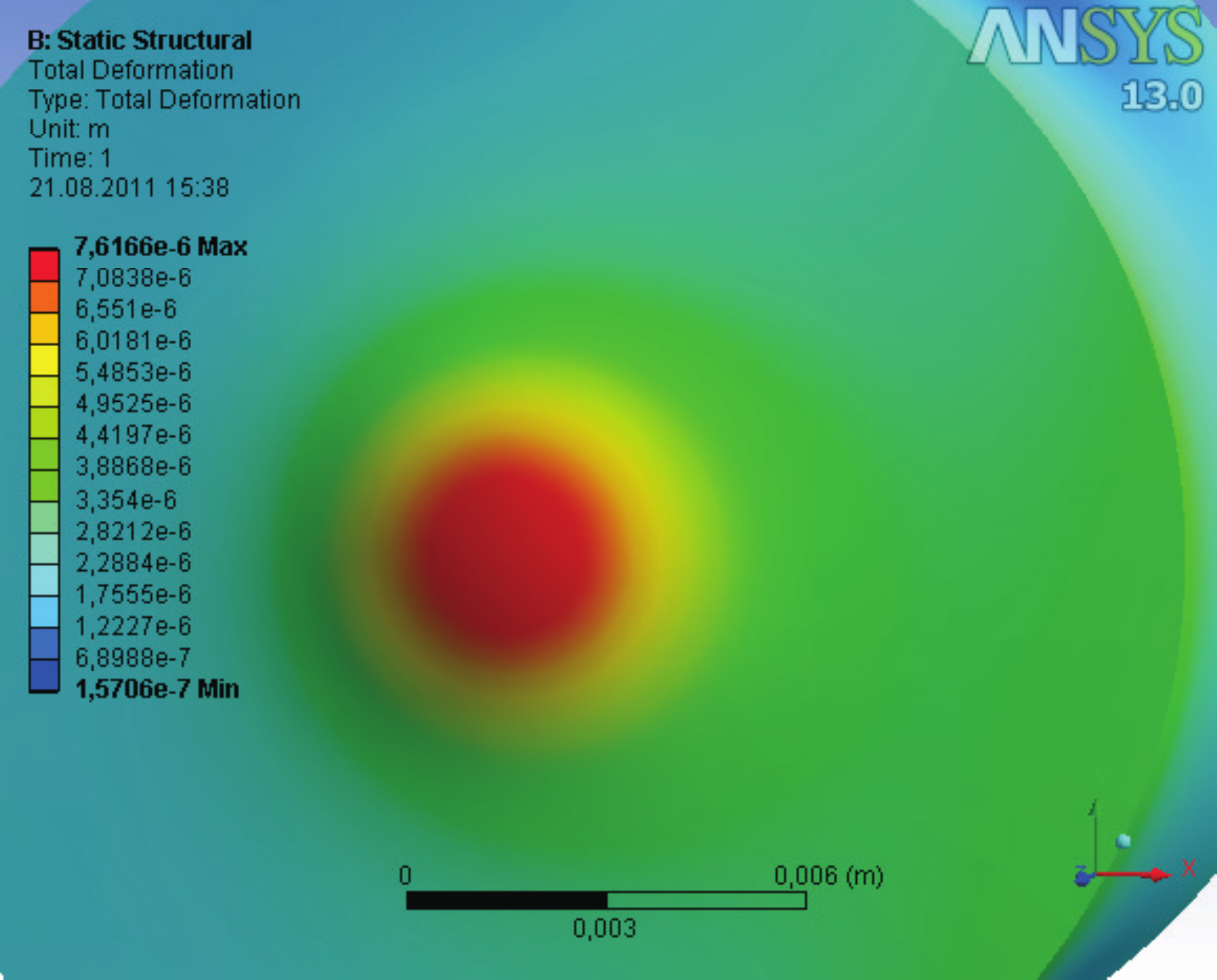}
\includegraphics[height=2.35in,keepaspectratio=true]{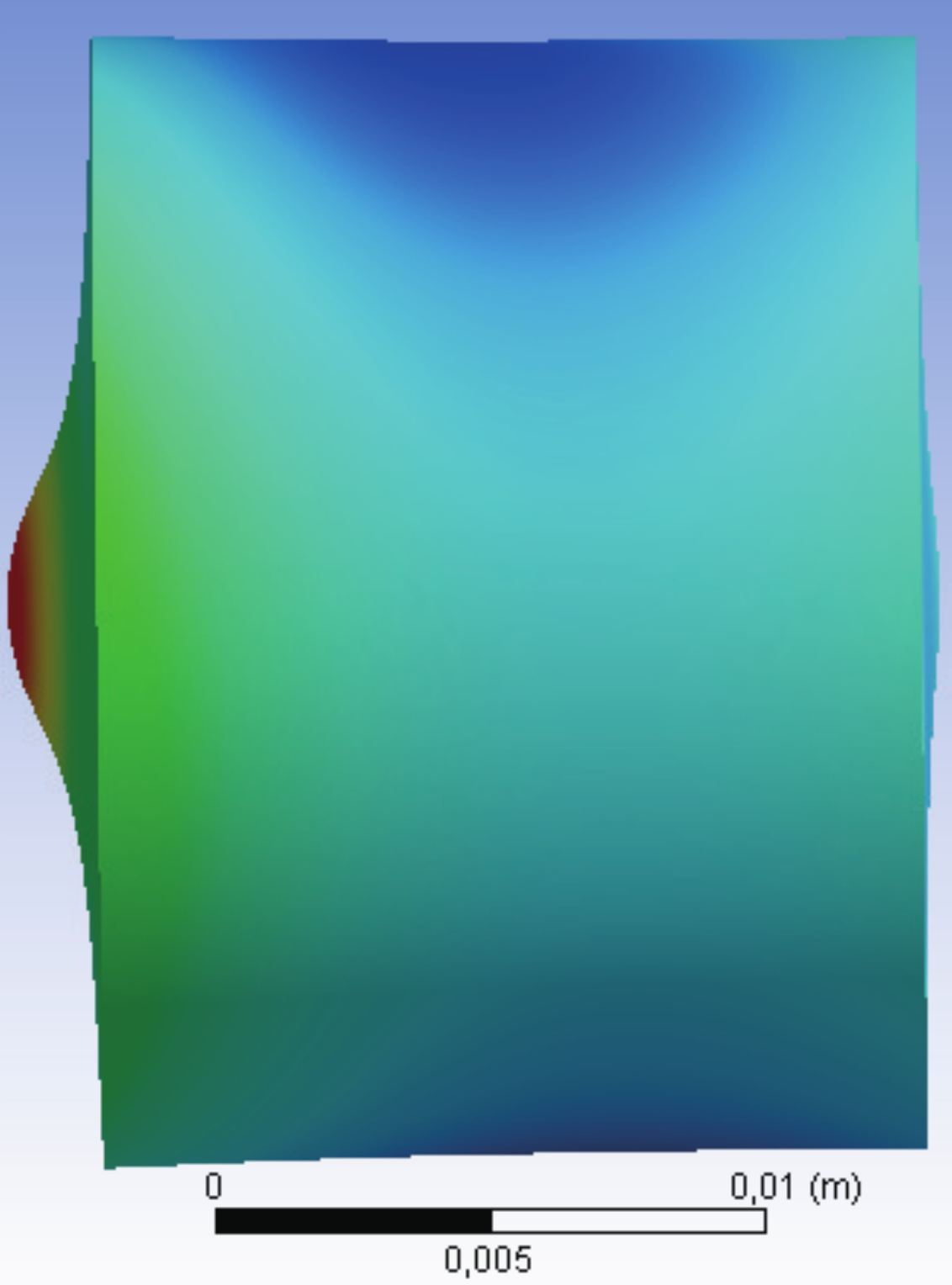}
\end{center}
\caption{Total deformation of the target after 100 bunches (back
view -- left, side view -- right).}
\label{au:fig-Deform}
\end{figure}

\begin{figure}
\begin{center}
\includegraphics[height=2.9in,keepaspectratio=true]{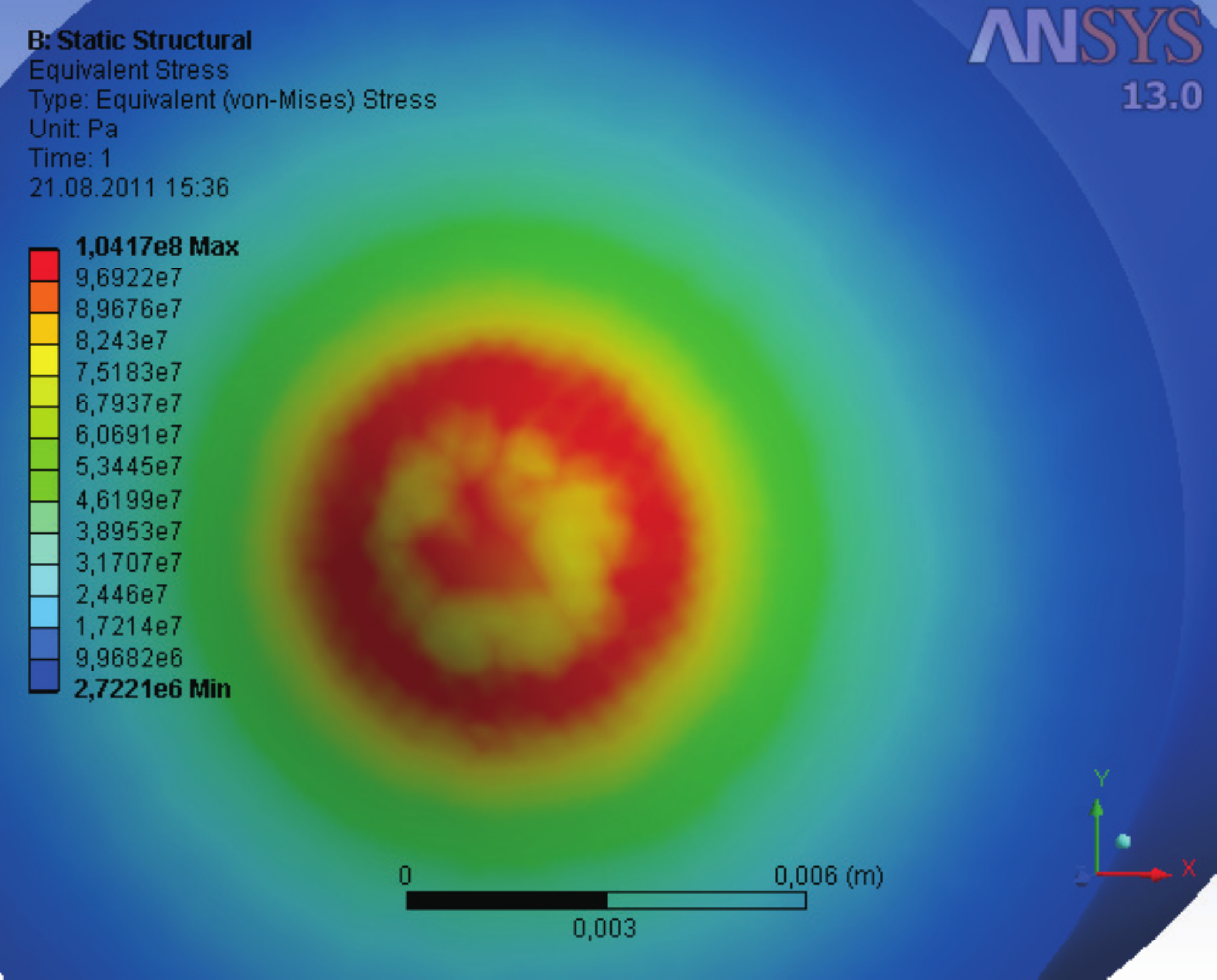}
\end{center}
\caption{Von-Mises stress after 100 bunches (ANSYS static structural
model).}
\label{au:fig-Stress}
\end{figure}

As a first step, a statical ANSYS model of the target material
response to the heat load (see, Figures \ref{au:fig-Edep-fluka} and
\ref{au:fig-Temperature-backside}) has been applied. The total
deformation and equivalent von-Mises stress are shown in Figs. \ref
{au:fig-Deform} and \ref{au:fig-Stress}. The maximum of equivalent
stress is about 100 MPa on the back side of the target in the
circular area around the photon beam axis with a radius of approx. 2
mm. This stress is about 12\% of the tensile yield strength for
titanium alloy (the properties of Ti6Al4V alloy, grade 5 can be
found, for example, in Ref. \cite{ASM}).

\section{Evolution of Thermal Stress in Time}

To simulate the time evolution of thermal stress in the positron
source target, the target movement has been analyzed more accurately
and ANSYS transient (explicit) model of deformation and stress has
been used.

\begin{figure}
\begin{center}
\includegraphics[height=1.8in,keepaspectratio=true]{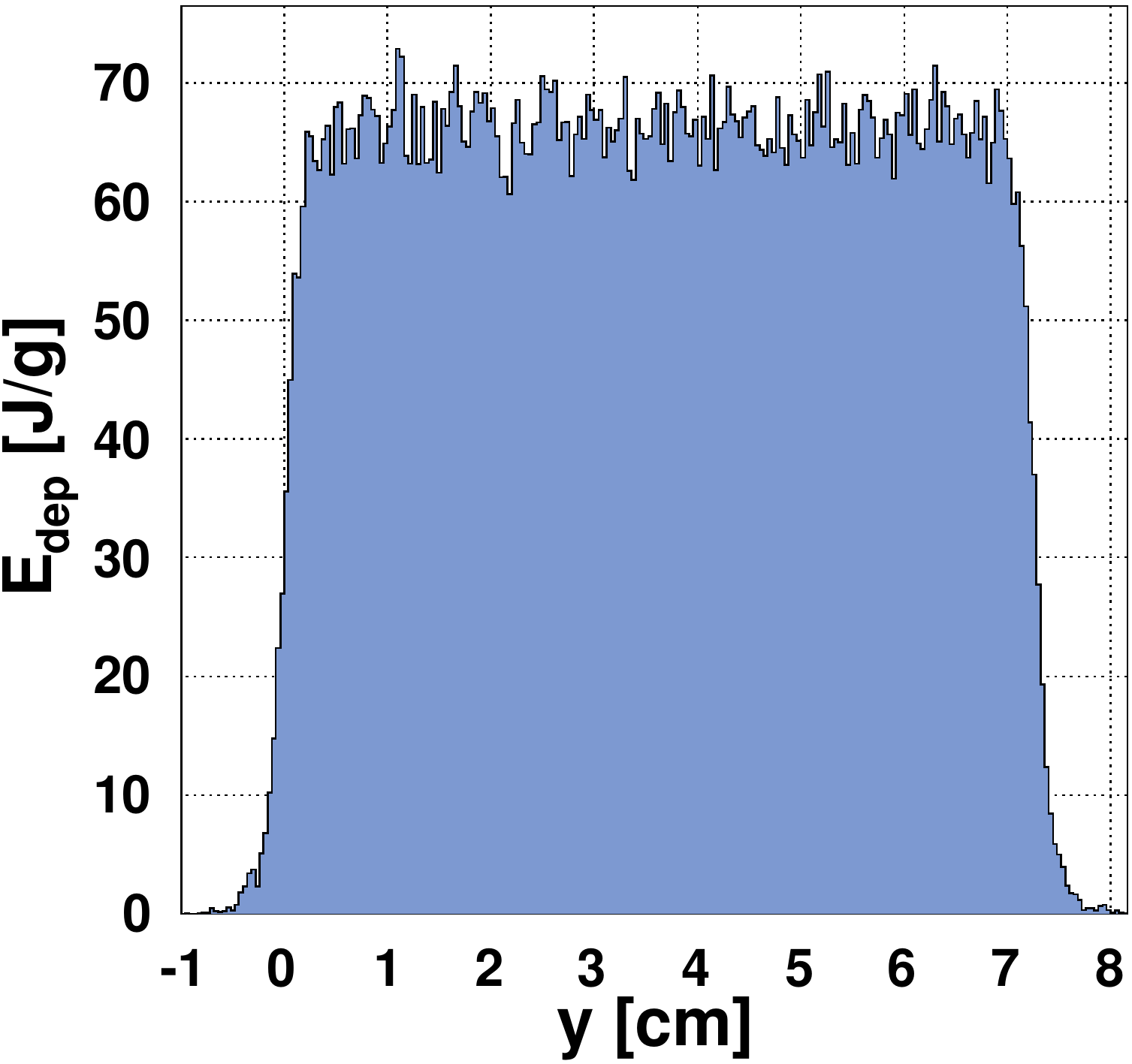} $~~~~~$
\includegraphics[height=1.8in,keepaspectratio=true]{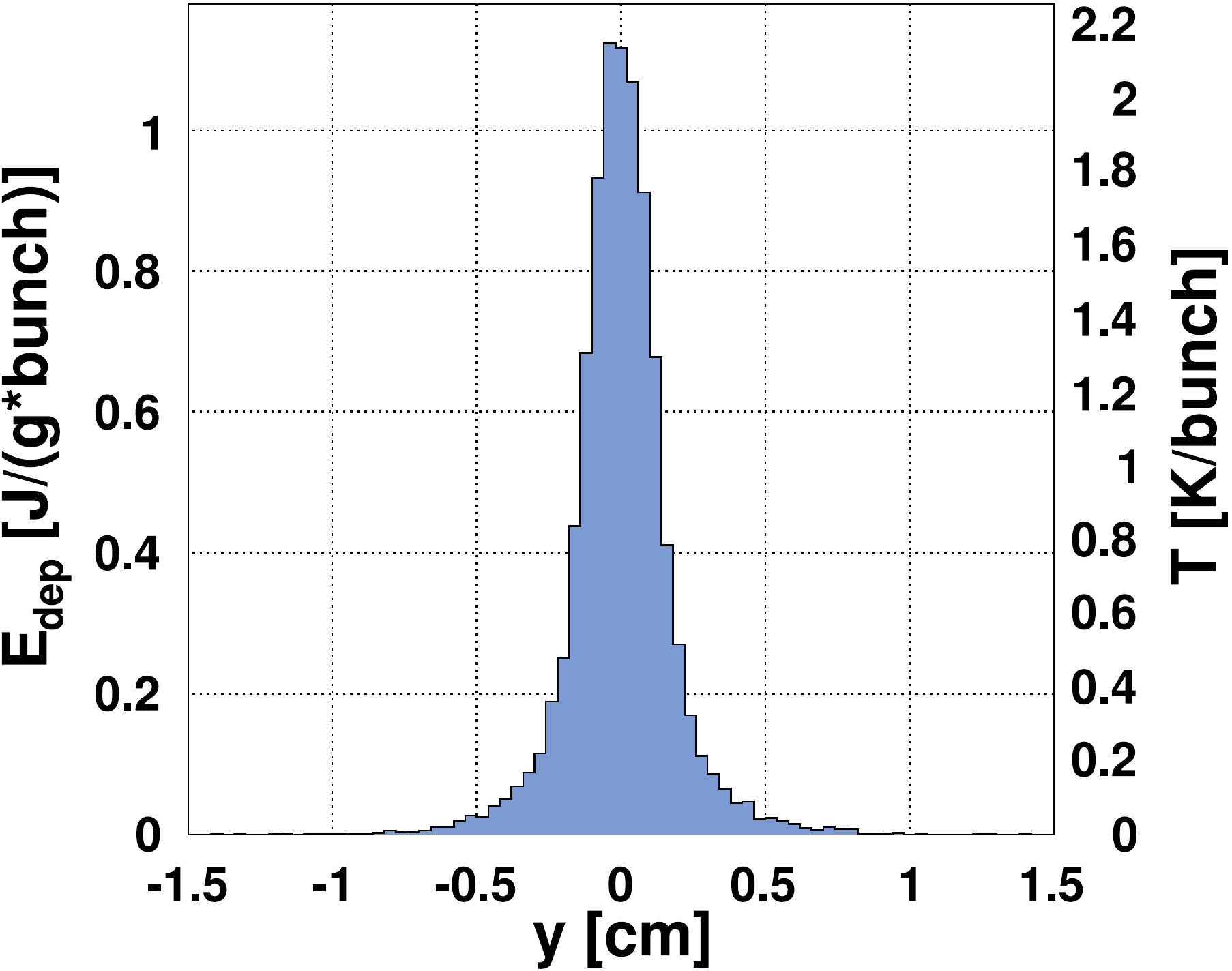}
\end{center}
\caption{Energy deposition in a target rotated with 100 m/s: left
-- after one pulse, right -- after one bunch.}
\label{au:fig-Edep-rot}
\end{figure}

We consider the tangential velocity of the target rim (1~meter in
diameter) of 100 meters per second as velocity in $y$ direction in a
Cartesian system. The energy deposited after one pulse (1312 bunches
with 554 ns bunch separation) as function of $y$ coordinate is shown
in Fig. \ref {au:fig-Edep-rot}. This Figure shows also the energy
deposited by a single bunch and the corresponding temperature rise.
Both profiles on Fig.~ \ref{au:fig-Edep-rot} are plotted for highest
energy deposition: in the $z$-direction -- close to the target back
side and in the middle of bunch(es) -- in $x$-direction. The bunch
overlapping factor is defined as the ratio of the maximal deposited
energy after a complete bunch train with respect to the maximum
after just one bunch. This factor for the nominal SB2009 source
parameters is about 59.

\begin{figure}
\begin{center}
\includegraphics[height=2.8in,keepaspectratio=true]{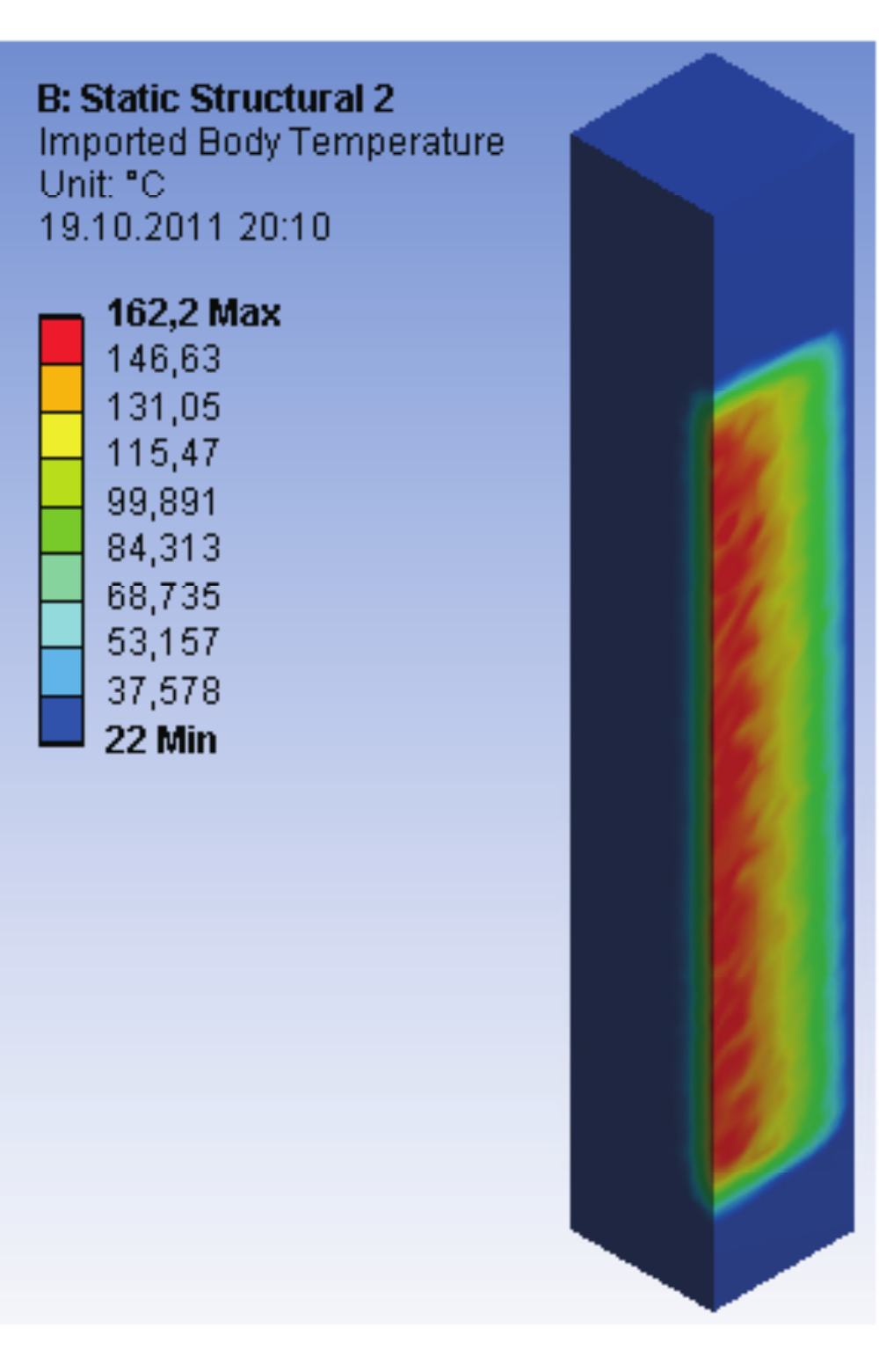}
$~$
\includegraphics[height=2.8in,keepaspectratio=true]{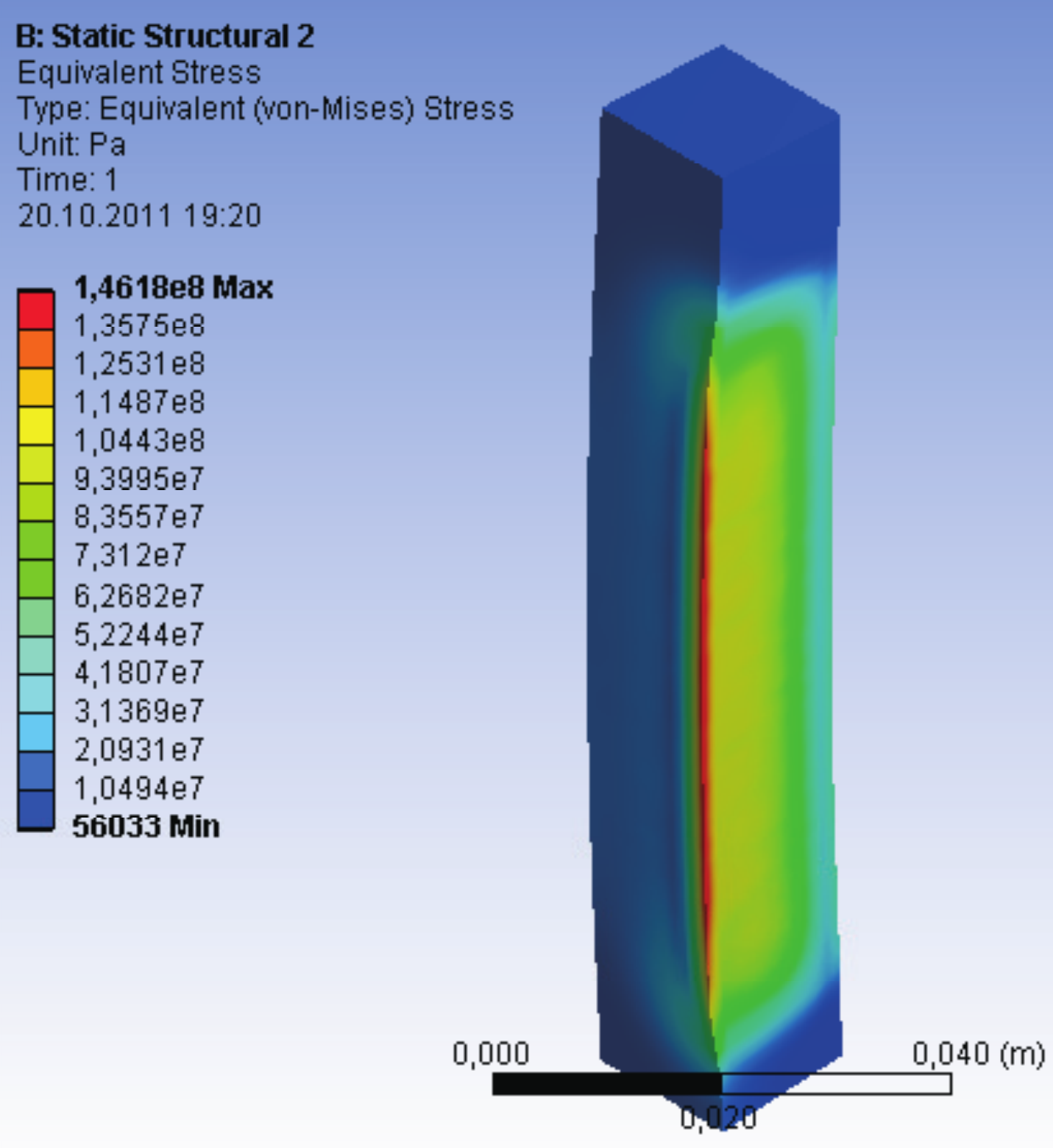}
\end{center}
\caption{Temperature distribution and induced equivalent stress in the
rotated target.}
\label{au:fig-TempStress-rot}
\end{figure}

Figure \ref{au:fig-TempStress-rot} shows the temperature and
equivalent stress in the ``rotated'' target. The target has been cut
in the middle plane in order to show the distributions inside the
target. The static ANSYS model for the equivalent stress after one
bunch train does not take into account thermal diffusion and thus
overestimates the stress induced in the target.

\begin{figure}
\begin{center}
\includegraphics[height=1.4in,keepaspectratio=true]{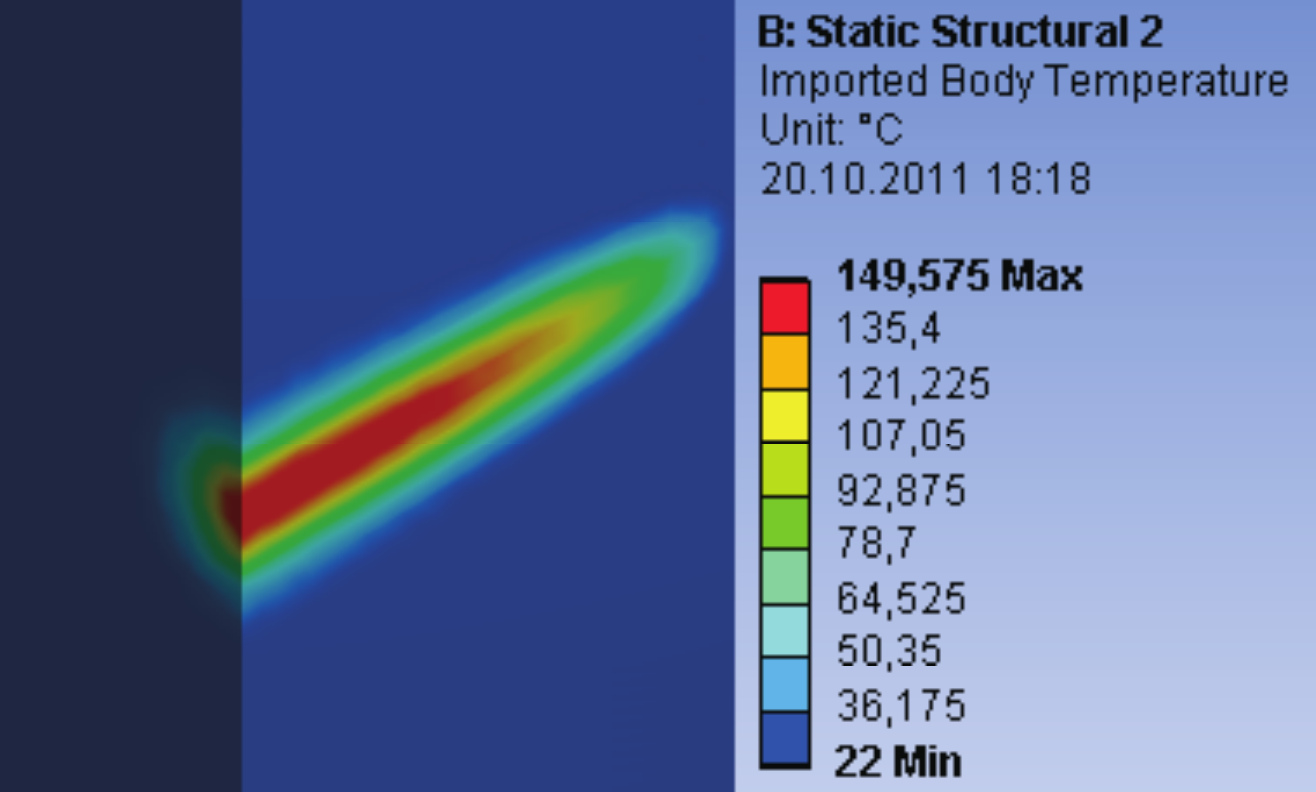}
\end{center}
\caption{Temperature distribution after 59 bunches.}
\label{au:fig-Temperature59}
\end{figure}

To reduce the effect of thermal diffusion on the stress and to study
the time-dependent dynamic effects, another model has been used. In
this model the temperature distribution after one single bunch has
been scaled with the above-mentioned bunch overlapping factor. The
cylindrical geometry of the target has been chosen to keep the
symmetry of the model and to reduce the computing time. Figure~\ref
{au:fig-Temperature59} shows the temperature distribution after ``59
bunches''.

\begin{figure}
\begin{center}
\includegraphics[height=2.3in,keepaspectratio=true]{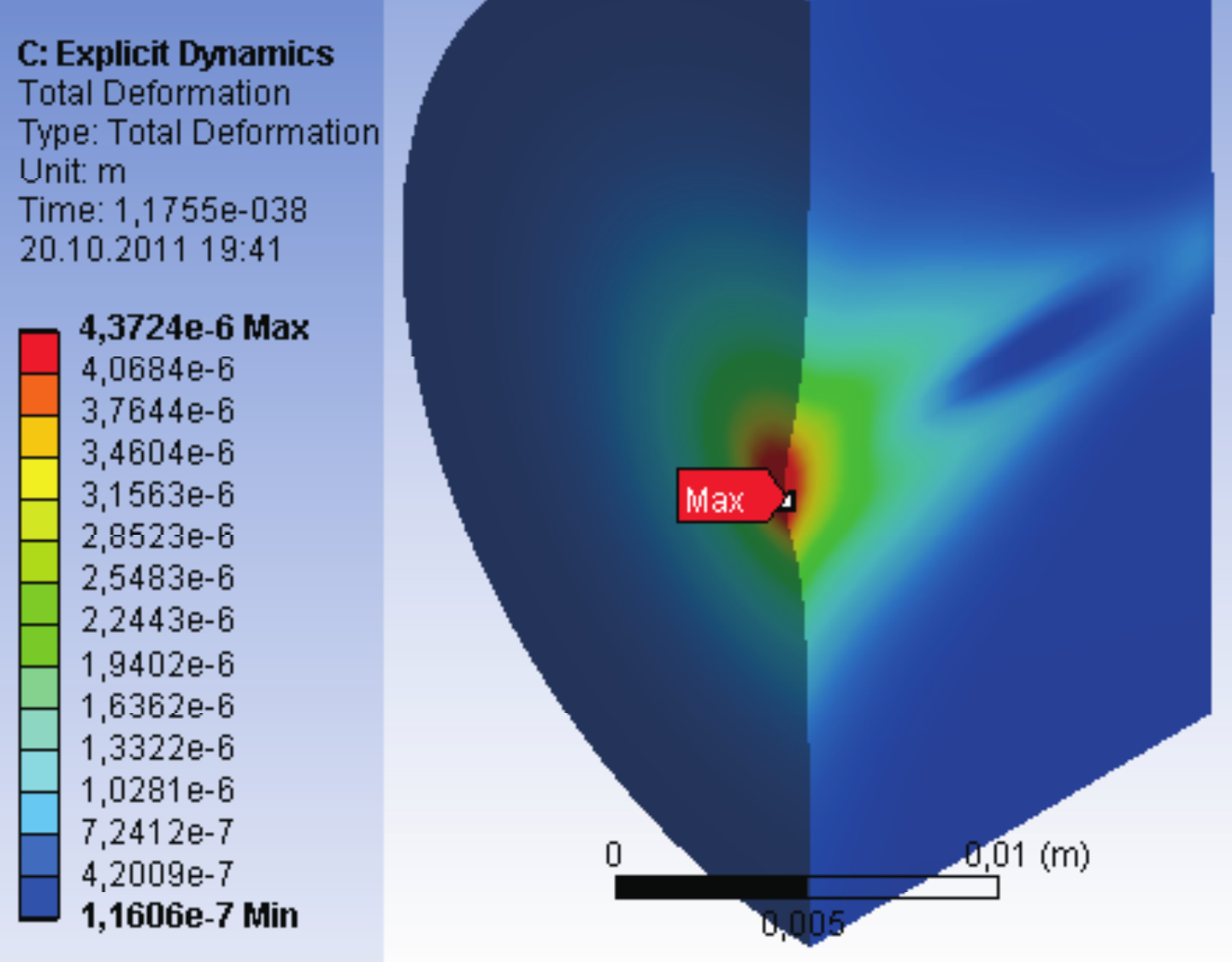}
\end{center}
\caption{Distribution of total deformation after 59 bunches.}
\label{au:fig-Deformation59}
\end{figure}

\begin{figure}
\begin{center}
\includegraphics[height=2.0in,keepaspectratio=true]{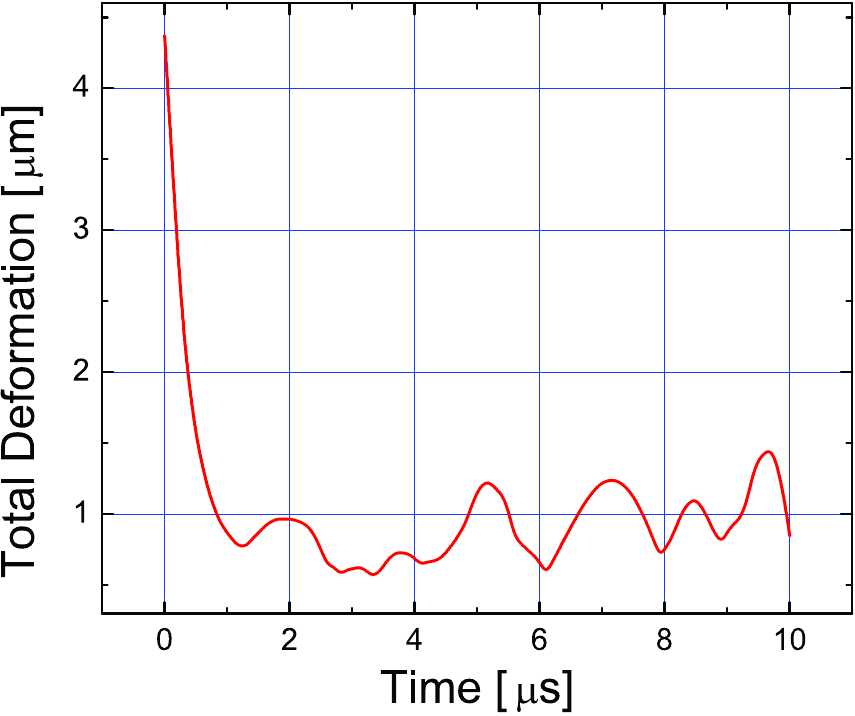}
\end{center}
\caption{Time evolution of maximal total deformation after 59 bunches.}
\label{au:fig-Deformation59-vs-time}
\end{figure}

The total deformation after 59 bunches is plotted in Fig.~\ref
{au:fig-Deformation59} and the evolution in time of maximal
deformation is shown in Fig.~\ref{au:fig-Deformation59-vs-time}. The
starting time (0 sec.) corresponds to the end of the pulse. The
reflections from the target surfaces and interference of the waves
result in the series of maxima at the level about 25\% of the
initial deformation.

\begin{figure}
\begin{center}
\includegraphics[height=2.0in,keepaspectratio=true]{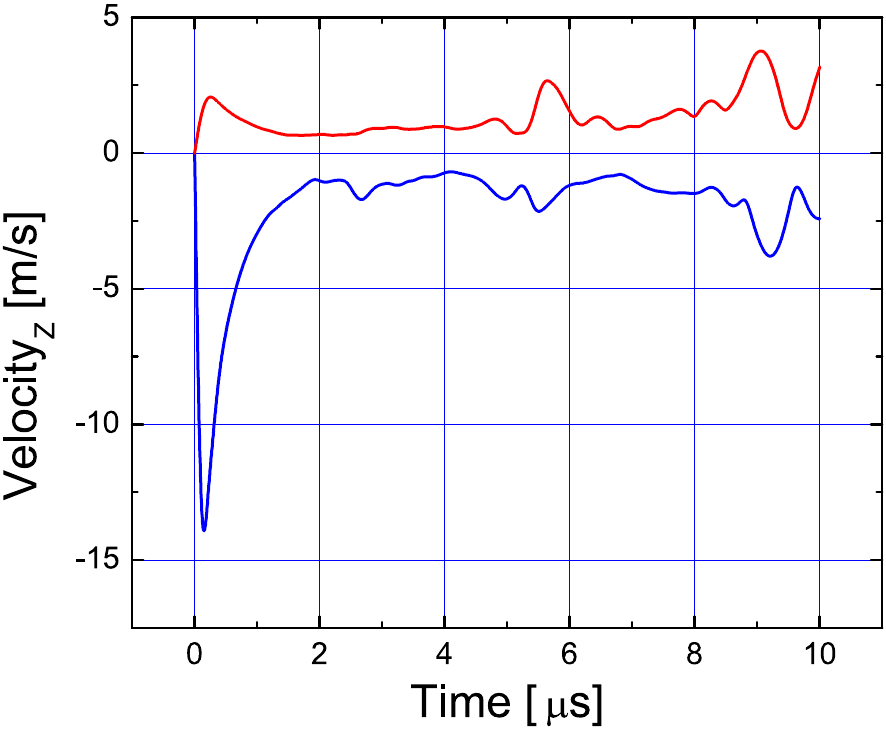}
\end{center}
\caption{Time evolution of maximal and minimal $z$-component of
velocity after 59~bunches.}
\label{au:fig-VelocityZ59-vs-time}
\end{figure}

\begin{figure}
\begin{center}
\includegraphics[height=2.3in,keepaspectratio=true]{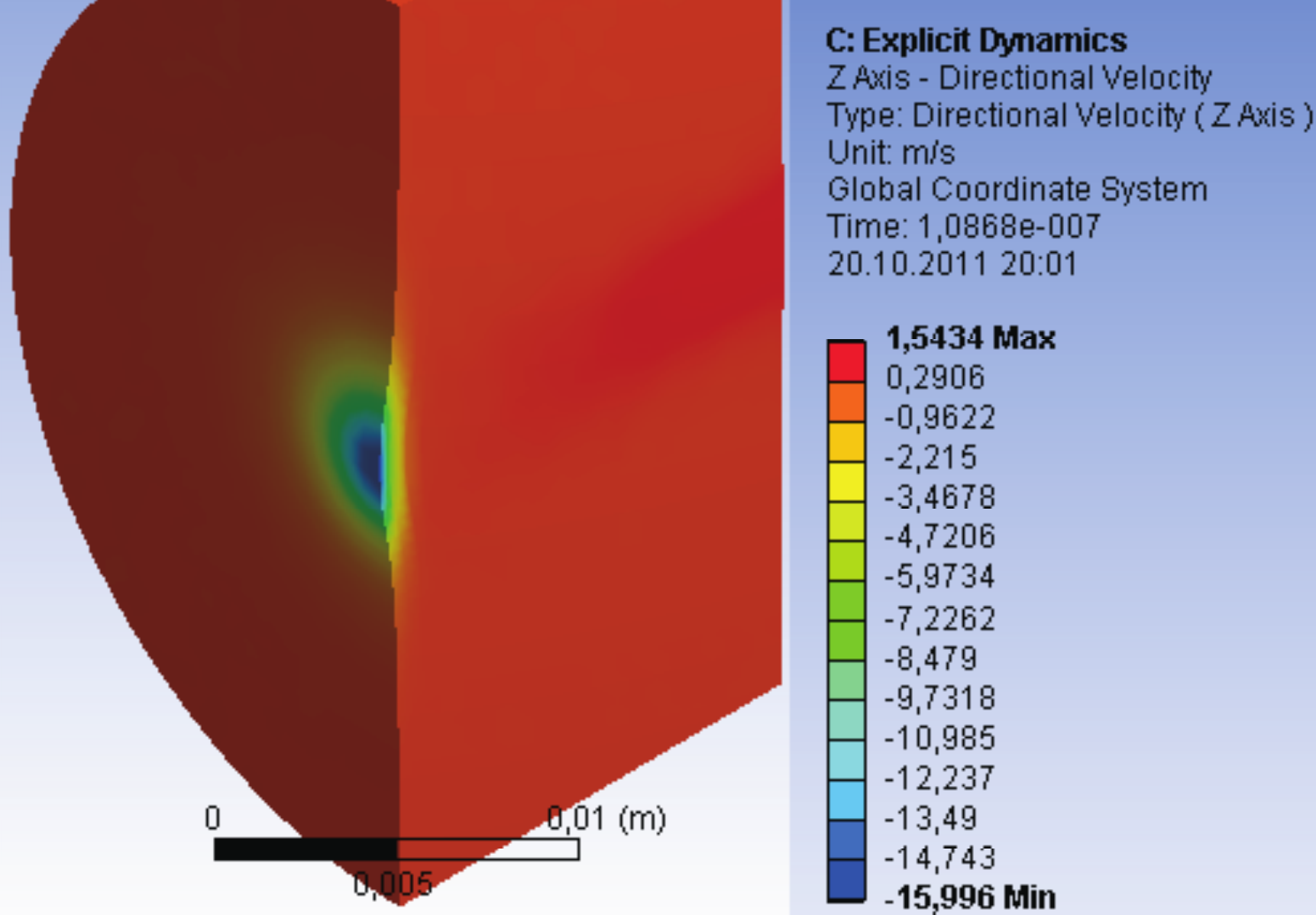}
\end{center}
\caption{Velocity along $z$-axis after 59 bunches and 0.1 $\mu$s delay.}
\label{au:fig-VelocityZ59}
\end{figure}

\begin{figure}
\begin{center}
\includegraphics[height=2.0in,keepaspectratio=true]{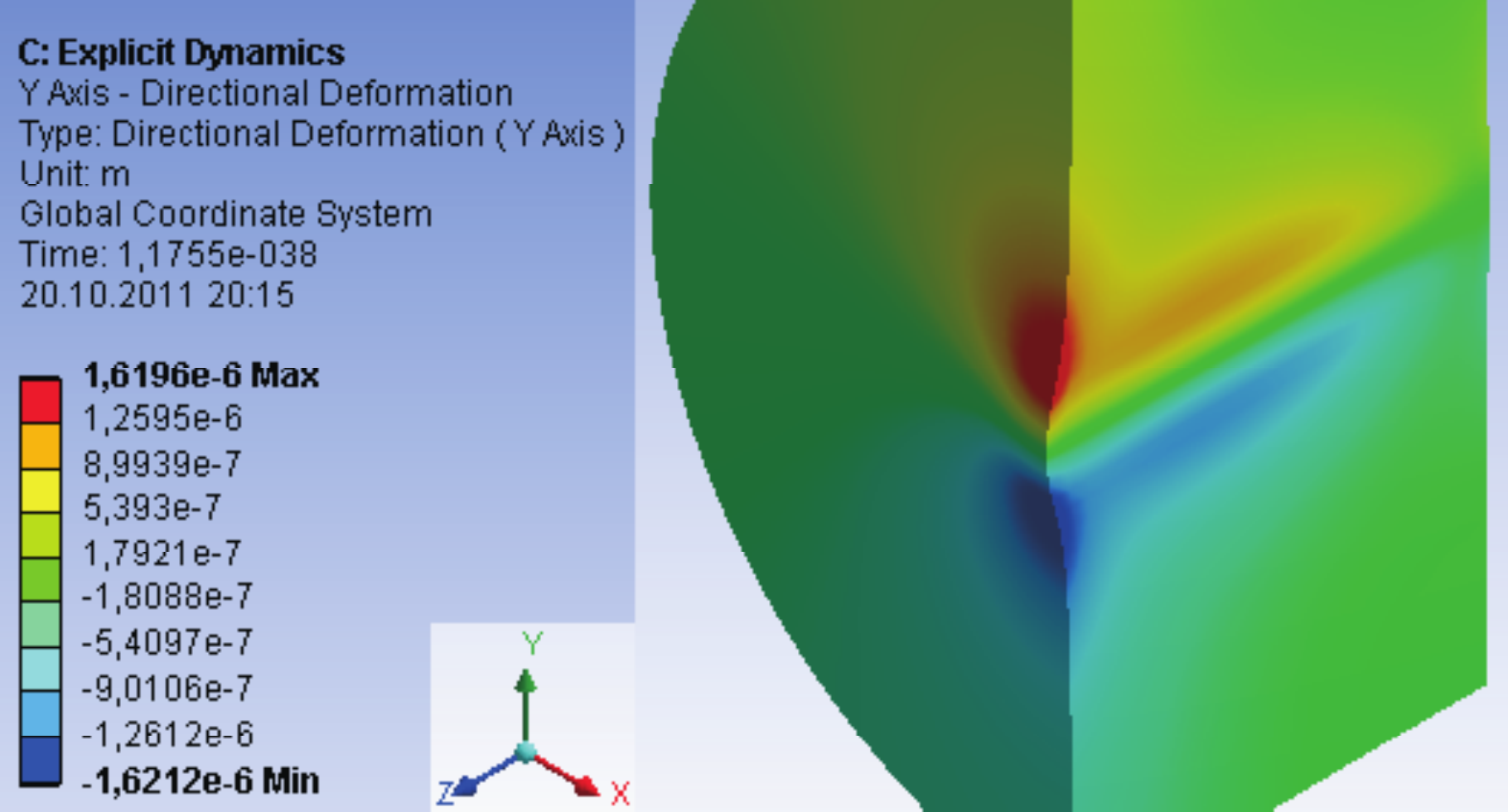}
\end{center}
\caption{Direction ($y$-axis) deformation after 59 bunches.}
\label{au:fig-DeformationY59}
\end{figure}

\begin{figure}
\begin{center}
\includegraphics[height=2.0in,keepaspectratio=true]{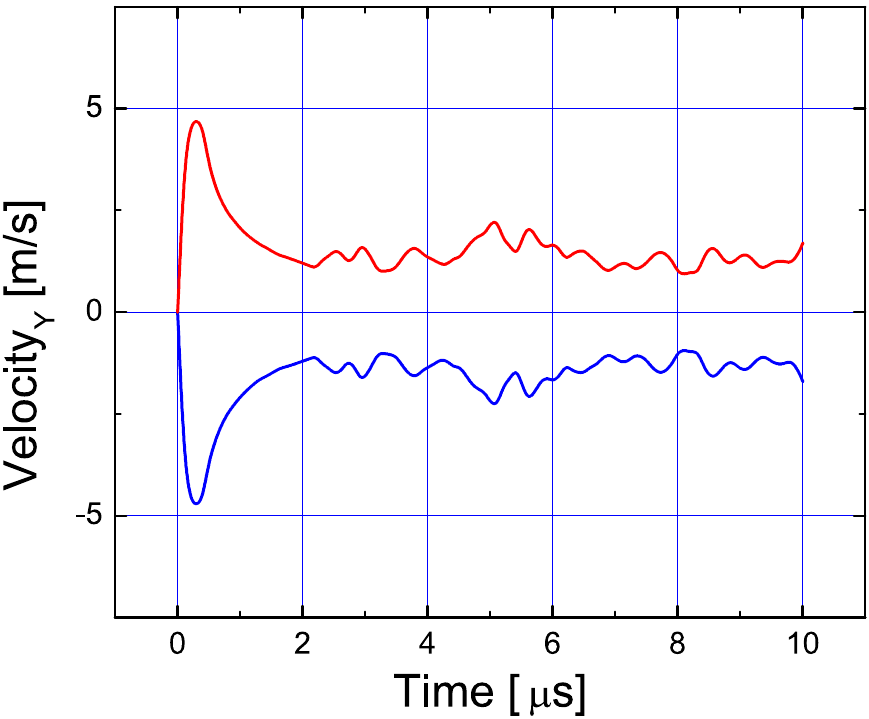}
\end{center}
\caption{Time evolution of maximal and minimal $y$-component of
velocity after 59~bunches.}
\label{au:fig-VelocityY59-vs-time}
\end{figure}

The deformation transverse to the beam axis (radial deformation)
contributes only minor (about one third) to the total deformation.
The time dependence of the dominating longitudinal ($z$-component)
velocity is presented  in Fig.~\ref{au:fig-VelocityZ59-vs-time}
showing the positive velocity directed out of the target and
negative velocity. Figure~\ref{au:fig-VelocityZ59} shows a snapshot
of the time evolution for the $v_{z}$-distribution after one pulse
and with 0.1 $\mu$s delay at the moment when the negative velocity
has reached the maximum. The $y$-component of deformation and
velocity are also shown in Figs.~\ref{au:fig-DeformationY59} and
\ref{au:fig-VelocityY59-vs-time}. Because of geometry and beam
symmetry, the deformation, maximal and minimal $v_{y}$-dependencies
on time are symmetrical (mirrored) too. It has to be noted that the
transient effects during the pulse were not considered and thus the
all velocities are starting from zero level at the end of the beam
pulse.

\begin{figure}
\begin{center}
\includegraphics[height=2.0in,keepaspectratio=true]{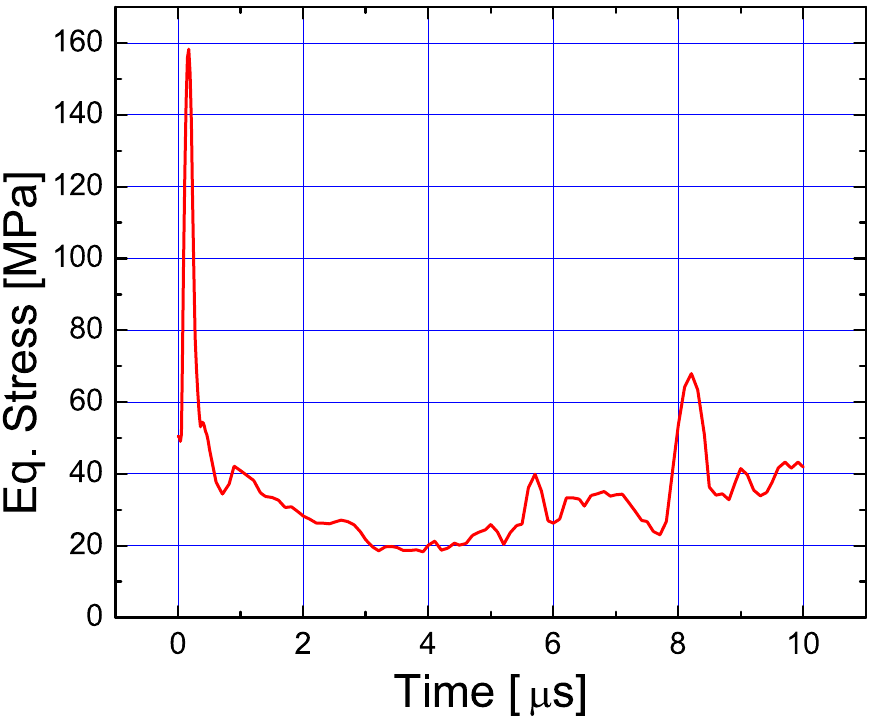}
\end{center}
\caption{Time evolution of maximal equivalent stress after 59~bunches.}
\label{au:fig-Stress59-vs-time}
\end{figure}

\begin{figure}
\begin{center}
\includegraphics[height=1.67in,keepaspectratio=true]{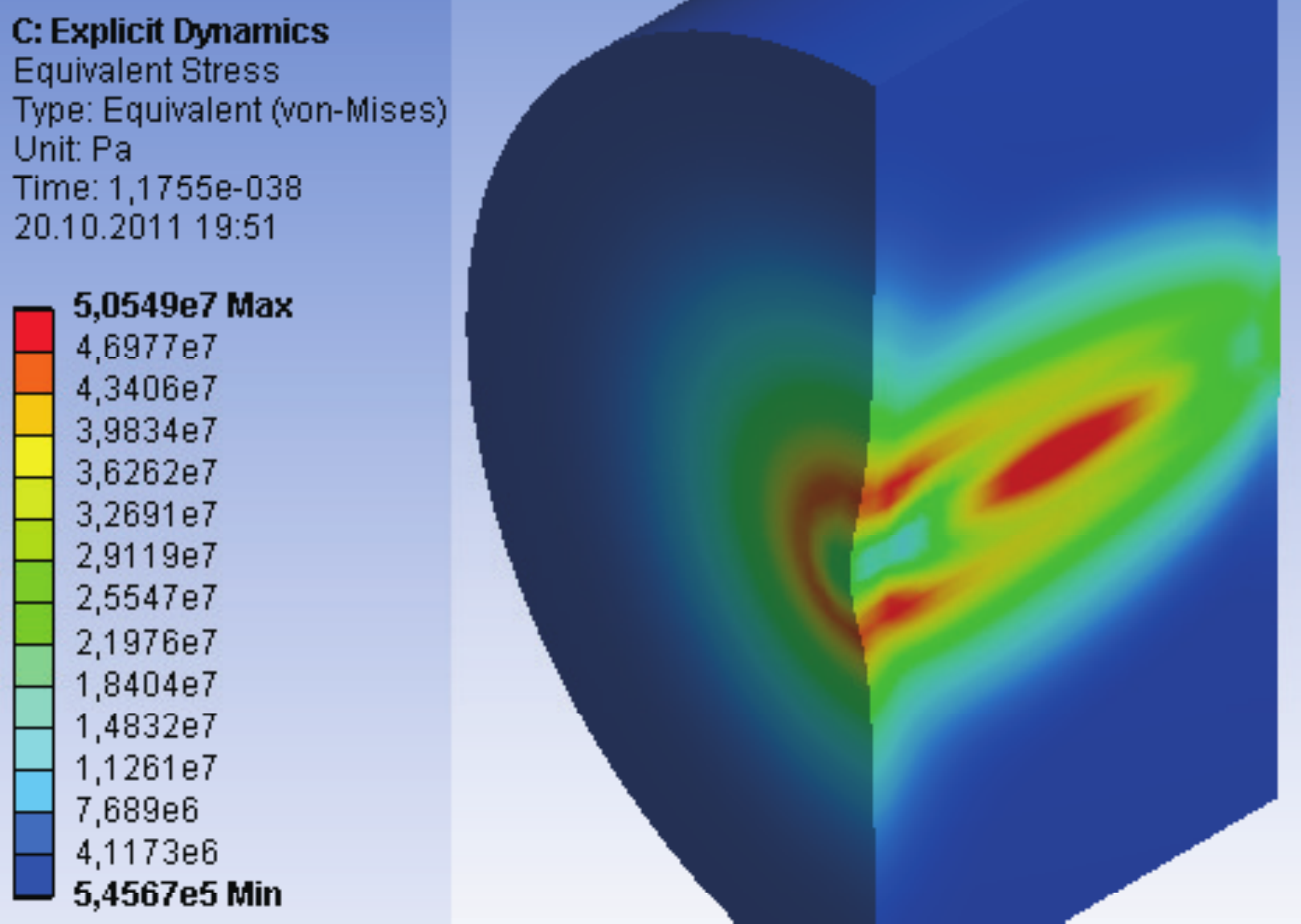}
\includegraphics[height=1.67in,keepaspectratio=true]{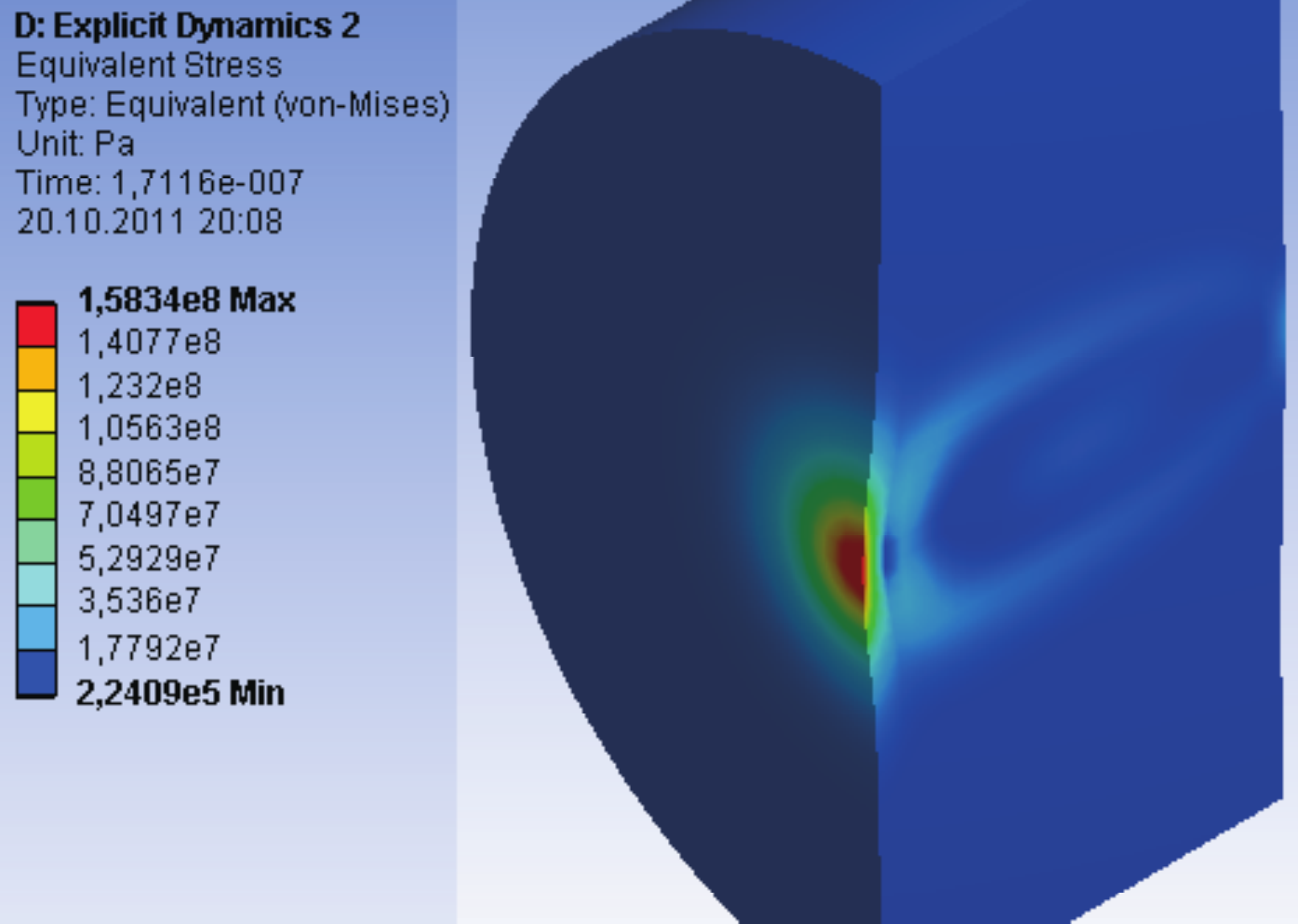}
\end{center}
\caption{Equivalent stress after 59 bunches (left) and additional
0.1~$\mu$s delay (right).}
\label{au:fig-Stress59}
\end{figure}

The time evolution of the maximal equivalent (von-Mises) stress in
the target is plotted in Fig.~\ref{au:fig-Stress59-vs-time}. The
stress distributions after one beam pulse and additional 0.1 $\mu$s
delay are shown in Fig.~\ref{au:fig-Stress59}. The peak stress value
is about 160 MPa which corresponds to 18\% of tensile yield
strength. This level of stress can be considered as acceptable.

\section*{Summary}
The energy deposition in the ILC positron source target has been
simulated in FLUKA for the SB2009 set of parameters. The peak energy
density in the rotated titanium-alloy target  is about 120~J/g for
the conservative choice of a magnetic focusing device (quarter-wave
transformer) and 250~GeV electron beam energy. The different
simplified (static and transient) ANSYS models have been used to
estimate the thermal stress induced by fast temperature rise and
thermal expansion of the target. The peak stress is about 160 MPa.
It is less then 20\% of tensile yield strength. Such stress will not
damage the target.

\section*{Outlook}
In the future, also the cooling of the target has to be added in
model. The procedure used so far, in which the deposited energy is
converted into temperature, has to be eliminated and the direct
import of the heat source into ANSYS can additionally improve the
accuracy of stress estimations. The thermal and structural effects
in the target have to be also simulated taking into account the time
structure of the bunch train.

\section*{Acknowledgments}
We would like to thank the organizers and the host of POSIPOL 2011 for this
fruitful and encouraging workshop and for hospitality.

Work supported by the German Federal Ministry of Education and Research,
Joint Project R\&D Accelerator ``Spin Management'', contract No.~05H10GUE.

\end{document}